\documentclass[prd,preprint,superscriptaddress]{revtex4}
\usepackage{amssymb}
\usepackage{revsymb}
\usepackage{graphicx}

\begin{document}

\title{Relic gravitational waves and the generalized second law}
\author{Germ\'{a}n Izquierdo \footnote{
E-mail address: german.izquierdo@uab.es}}
\affiliation{Departamento de F\'{\i}sica, Universidad Aut\'{o}noma de Barcelona, 08193
Bellaterra (Barcelona), Spain}
\author{Diego Pav\'{o}n\footnote{
E-mail address: diego.pavon@uab.es}}
\affiliation{Departamento de F\'{\i}sica, Universidad Aut\'{o}noma de Barcelona, 08193
Bellaterra (Barcelona), Spain}

\begin{abstract}
The generalized second law of gravitational thermodynamics is
applied to the present era of accelerated expansion of the
Universe. In spite of the fact that the entropy of matter and
relic gravitational waves inside the event horizon diminish, the
mentioned law is fulfilled provided the expression for the entropy
density of the gravitational waves satisfies a certain condition.
\end{abstract}

\maketitle
\section{Introduction}
As is well--known, when the Universe makes a transition from a
stage of expansion dominated by a given energy source to the next
(e.g., inflation--radiation era, radiation--matter era,
matter--dark energy era), relic gravitational waves (RGWs), i.e.,
the gravitational waves generated  from the quantum vacuum, suffer
amplification \cite{leonid,aas,allen,maia}. Likewise, their
wavelength get stretched as the Universe expands. For a wave mode
$k$ to fit  in the Hubble horizon ($H^{-1}$) it must fulfill the
obvious condition $k > aH$ (where $a$ is the scale factor of the
Robertson-Walker metric). During decelerated eras of cosmic
expansion the quantity $aH$ decreases and gravitational waves
continuously enter the horizon. By contrast, in accelerated eras
-like the present one, see e.g.  \cite{accel}- $aH$ increases
whereby gravitational waves continuously leave the horizon. Those
most recently created are the first to exit, those created in the
inflationary era are the last to exit.

On the other hand, gravitational waves whose wavelength exceeds
the horizon size do not contribute to the energy density in the
horizon \cite{allen,maia} and so they do not add to the entropy
inside the horizon. Therefore, in the current accelerated era of
expansion the gravitational wave entropy in the horizon is
steadily diminishing and, as we shall see, the entropy of matter
is decreasing as well. As for the entropy of the dark energy field
responsible for the acceleration we can only say that it is likely
zero (as is the case of the cosmological constant) or
undetermined. There still remains the entropy of the horizon
itself. Properly speaking, the entropy of the event horizon is
given by  $S_{H}= (k_{B}/4) {\cal A}/\ell_{Pl}^{2}$ -where ${\cal
A}$ is the area of the horizon, $k_{B}$ the Boltzmann constant,
and $\ell_{Pl}$ the Planck length-, see Ref. \cite{GH}.

By extending the ``generalized second law" (GSL) of black hole spacetimes \cite{bhth}
to cosmological settings, several authors have considered the interplay between
ordinary entropy and the entropy associated to cosmic event horizons \cite{paul,gsl}.
According to this law, the entropy of matter and/or radiation  within the horizon
plus the entropy of the event horizon cannot diminish with time. In this Brief
Report we study, with the help of the GSL, the entropy budget of matter,
gravitational waves and event horizon in the current accelerated stage of
cosmic expansion. As it turns out, the GSL is fulfilled provided a
proporcionality constant entering the expression of the gravitational
wave entropy satisfies a certain upper bound.

Section II gives the power spectrum of the relic gravitational waves at the
beginning of the present era of accelerated expansion. Section III considers
the evolution of the entropy of the gravitational waves during that era. Finally,
in section IV application of the GSL is seen to imply an upper bound on
the expression of the entropy density of gravitational waves.

\section{The relic gravitational waves in the dark energy era}
\begin{figure}[tbp]
\includegraphics*[angle=0,scale=0.7]{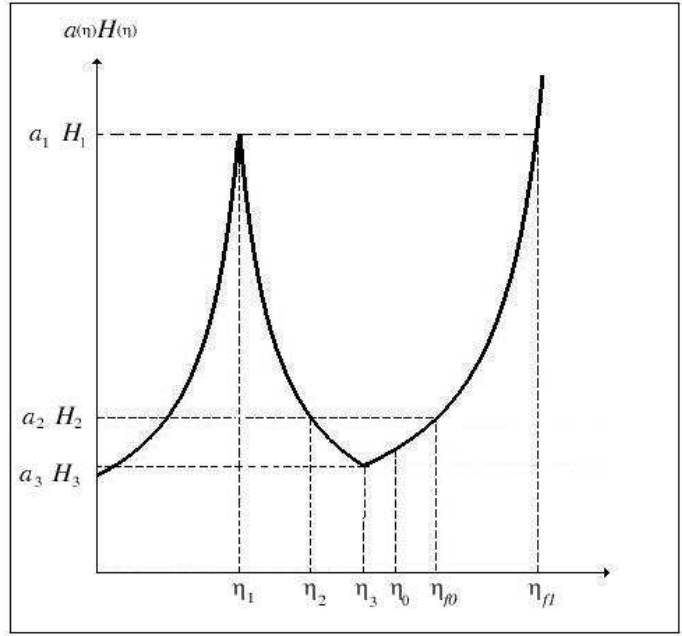}
\caption{{}Evolution of $a(\protect\eta )H(\protect\eta )$ in a universe
with scale factor given by Eq. (\protect\ref{sclfac}).}
\label{aH}
\end{figure}

The current era of cosmic acceleration is believed to be dominated by some sort
of energy (generically called ``dark energy") characterized by violating the
strong energy condition \cite{reviews}. In a recent paper we studied the
spectrum and energy density of RGWs \cite{lategw}. We assumed the
present era was successively preceded  by an inflationary era, a radiation
dominated era and a matter dominated era. The dependence of the scale
factor on the conformal time in these eras is given by
\\
\begin{equation}
a(\eta )=\left\{
\begin{array}{c}
-\frac{1}{H_{1}\eta }\qquad (-\infty <\eta <\eta _{1}<0),\qquad \text{De
Sitter era} \\
\frac{1}{H_{1}\eta _{1}^{2}}(\eta -2\eta _{1})\qquad (\eta _{1}<\eta <\eta
_{2}),\qquad \text{radiation era} \\
\frac{1}{4H_{1}\eta _{1}^{2}}\frac{(\eta +\eta _{2}-4\eta _{1})^{2}}{\eta
_{2}-2\eta _{1}}{\qquad }(\eta _{2}<\eta <\eta _{3}),\qquad \text{dust era}
\\
\left( \frac{l}{2}\right) ^{-l}\frac{(\eta _{3}+\eta _{2}-4\eta _{1})^{2-l}}{%
4H_{1}\eta _{1}^{2}(\eta _{2}-2\eta _{1})}\left( \eta _{l}\right) ^{l}{%
\qquad }(\eta _{3}<\eta ),\qquad \text{dark energy era}%
\end{array}%
\right.  \label{sclfac}
\end{equation}%
where $l < -1$, $\eta _{l}=\eta +\frac{l}{2}\left[ -\textstyle{2\over{l+1}}
\, \eta_{3}+\eta _{2}-4\eta _{1}\right] $. The subindexes $1,2,3$ correspond
to sudden transitions from De Sitter era to radiation era, from radiation to
dust era, and from dust era to dark energy era, respectively, $H_{i}$ is the
Hubble factor at the instant $\eta =\eta _{i}$. The present time $\eta _{0}$
lies in the dark energy era. The Hubble factor during the dark energy era
obeys
\\
\begin{equation}
H(\eta )=\left( \frac{a_{3}}{a(\eta )}\right) ^{1+\frac{1}{l}}H_{3}.
\label{h}
\end{equation}
\\
The evolution of the quantity $aH$ in terms of the conformal
time $\eta$ is sketched in Fig. \ref{aH}.
\\

The modes' solution to the gravitational wave equation during the
De Sitter era are related with those of the final dark energy era
by a Bogoliubov transformation with coefficients $\alpha _{Tr}$
and $\beta _{Tr}$. Specifically, $\left\langle N_{\omega
}\right\rangle=\left\vert \beta _{Tr}\right\vert ^{2}$ gives the
number of RGWs created from the initial vacuum state, and from it
we get the power spectrum $P(\omega (\eta ))=\left( \hbar \omega ^{3}(\eta%
)/\pi ^{2}c^{3}\right) \left\langle N_{\omega }\right\rangle $, where $\omega$
denotes the frequency. At the beginning of the dark energy era,
$\eta = \eta_{3}$, the power spectrum (obtained in the adiabatic vacuum
approximation \cite{curved}) was \cite{lategw}
\\
\begin{equation}
P(\omega )\simeq \left\{
\begin{array}{l}
0\text{\qquad }\left( \omega (\eta )>2\pi (a_{1}/a(\eta ))H_{1}\right) , \\
\\
\frac{\hbar }{4\pi ^{2}c^{3}}\left( \frac{a_{1}}{a(\eta )}\right)
^{4}H_{1}^{4}\omega ^{-1}\text{\qquad }\left( 2\pi (a_{2}/a(\eta
))H_{2}<\omega (\eta )<2\pi (a_{1}/a(\eta ))H_{1}\right) , \\
\\
\frac{\hbar }{16\pi ^{2}c^{3}}\left( \frac{a(\eta )}{a_{2}}\right)
^{2}\left( \frac{a_{1}}{a(\eta )}\right) ^{8}H_{1}^{6}\omega ^{-3}\text{%
\qquad }\left( 2\pi H(\eta )<\omega (\eta )<2\pi (a_{2}/a(\eta
))H_{2}\right) .%
\end{array}%
\right.  \label{espectr}
\end{equation}%

During the radiation and dust eras $a(\eta )H(\eta )$ decreases with $\eta $ and
increases during the De Sitter and dark energy eras. Consequently,
RGWs are continuously leaving the Hubble radius during the
accelerated dark energy era \cite{lategw,Chiba}. At some instant $\eta_{f0}$,
defined by $a(\eta _{f0})H(\eta _{f0})=a_{2}H_{2}$, the third term in
(\ref{espectr}) ceases to contribute to the power spectrum since the
wavelengths of the corresponding RGWs exceed the size of the horizon.
Finally, at $\eta _{f1}$, defined by $a(\eta _{f1})H(\eta _{f1})=a_{1}H_{1}$,
all RGWs have their wavelength longer than the Hubble radius and the
power spectrum vanishes altogether.

\section{Entropy of the RGWs in the dark energy era}
There are different expressions in the literature for the entropy density of gravitational
waves -see e.g. \cite{gasperini, brandenberger,nesteruk}. All of them are based on the assumption
that the gravitational entropy is associated with the amount of RGWs
inside the horizon. We shall adopt the proposal of Nesteruk and Ottewill
\cite{nesteruk}, namely:
the gravitational entropy is proportional to the number of RGWs, i.e.,
\\
\begin{equation}
s_{g} = A\, n_{g}\, ,
\label{ng}
\end{equation}
\\
where $n_{g}$ is the number density of gravitational waves, and $A$ is an unknown
positive--definite constant of proportionality.

We are interested in the evolution of $n_{g}$ during the dark energy era.
The number density of RGWs created from the initial vacuum state is
\\
\begin{equation}
dn_{g}(\eta )=\left[ \frac{\omega ^{2}(\eta )}{2\pi ^{2}c^{3}}d\omega (\eta )%
\right] \left\langle N_{\omega }\right\rangle =\frac{P(\omega (\eta ))}{%
2\hbar \omega (\eta )}\, d\omega (\eta ),
\label{dng}
\end{equation}
\\
where the term in square brackets is the density of states. We can obtain $%
n_{g}(\eta )$ by inserting Eq. (\ref{espectr}) into Eq. (\ref{dng}) and
integrating over the frequency.

At the beginning of the dark energy era the RGWs number
density is
\\
\begin{equation}
n_{g}(\eta _{3})=n_{g}(\eta _{2})\left( \frac{a_{2}}{a_{3}}\right) ^{3}+%
\frac{1}{768\pi ^{5}c^{3}}\left( \frac{a_{1}}{a_{2}}\right) ^{2}\left( \frac{%
a_{2}}{a_{3}}\right) ^{3}H_{1}^{3}\left[ \left( \frac{a_{3}}{a_{2}}\right) ^{%
\frac{3}{2}}-1\right] ,
\end{equation}
\\
where
\\
\begin{equation}
n_{g}(\eta _{2})=\frac{1}{16\pi ^{3}c^{3}}\left( \frac{a_{1}}{a_{2}}\right)
^{3}H_{1}^{3}\left[ \frac{a_{2}}{a_{1}}-1\right]
\end{equation}
\\
is the number density at the transition radiation era--dust era.

For $\eta <\eta _{f0}$, i.e. $\frac{a(\eta )}{a_{3}}<\left( \frac{a_{3}}{%
a_{2}}\right) ^{-\frac{l}{2}}$, one has%
\begin{equation}
n_{g}(\eta _{3}<\eta <\eta _{f0})=n_{g}(\eta _{2})\left( \frac{a_{2}}{a(\eta
)}\right) ^{3}+\frac{1}{768\pi ^{5}c^{3}}\left( \frac{a_{1}}{a_{2}}\right)
^{2}\left( \frac{a_{2}}{a_{3}}\right) ^{3}H_{1}^{3}\left( \frac{a_{3}}{%
a(\eta )}\right) ^{3}\left[ \left( \frac{a_{3}}{a(\eta )}\right) ^{-\frac{3}{%
l}}\left( \frac{a_{3}}{a_{2}}\right) ^{\frac{3}{2}}-1\right] .
\label{dens1}
\end{equation}

The density number given by Eq. (\ref{dens1}) decreases with the scale factor
because of two effects: $(i)$ the evolution of the volume considered, which
grows as $a^{3}$, and $(ii)$ the exit of those waves whose wavelength becomes
longer than the Hubble radius, which appears in the term in square brackets.
As $\frac{a(\eta )}{a_{3}}$ approaches $\left( \frac{a_{3}}{a_{2}}\right) ^{-%
\frac{l}{2}}$, this term tends to zero. From this instant on, the
number density reads
\\
\begin{equation}
n_{g}(\eta _{f0}<\eta <\eta _{f1})=\frac{1}{16\pi ^{3}c^{3}}\left( \frac{%
a_{1}}{a_{3}}\right) ^{3}H_{1}^{3}\left( \frac{a_{3}}{a(\eta )}\right) ^{3}%
\left[ \left( \frac{a_{2}}{a_{1}}\right) ^{\frac{1}{2}}\left( \frac{a_{3}}{%
a_{1}}\right) ^{\frac{1}{2}}\left( \frac{a_{3}}{a(\eta )}\right) ^{-\frac{1}{%
l}}-1\right] .
\label{dens2}
\end{equation}
\\
As time goes on, the term in brackets tends to zero and at the instant
$\eta_{f1}$, $n_{g}$ vanishes.

Consequently, the entropy density, proportional to the number density,
decreases during the dark energy era not just because of the variation of
the volume considered but also because of the disappearance of the RGWs
from the Hubble volume.

The RGWs entropy inside the event horizon is
\\
\begin{equation}
S_{g}=\textstyle{4 \pi\over{3}} R_{H}^{3}s_{g}\, ,
\end{equation}
\\
where $R_{H} = a(t) \, \int_{t} ^{\infty}{ dt'/a(t')} \, $ is the radius of the event
horizon and $t$ the cosmic time. For the horizon to exist $R_{H}$ must not diverge.
For expansions of the general form  $a(\eta) \propto \eta^{l}$ (i.e.,
$a(t) \propto t^{n}$ with $n = l/(1+l)$ and $l < -1$), the horizon exists
and can be expressed as $R_{H} = -l c H^{-1}(\eta)$. Since it is of the order of
magnitude of the Hubble radius, $c H^{-1}$, we will neglect $-l$ and use both
terms interchangeably.

Making use of Eq. (\ref{h}) and Eqs. (\ref{dens1})-(\ref{dens2}) , we obtain
\\
\begin{eqnarray}
S_{g}(\eta _{3} &<&\eta <\eta _{f0})=\frac{4\pi c^{3}}{3}H_{3}^{-3}A\left(
\frac{a_{3}}{a(\eta )}\right) ^{-3/l}   \nonumber \\
&&\times \left\{ n_{g}(\eta _{2})\left( \frac{a_{2}}{a_{3}}\right) ^{3}+%
\frac{1}{768\pi ^{5}c^{3}}\left( \frac{a_{1}}{a_{2}}\right) ^{2}\left( \frac{%
a_{2}}{a_{3}}\right) ^{3}H_{1}^{3}\left[ \left( \frac{a_{3}}{a(\eta )}%
\right) ^{-\frac{3}{l}}\left( \frac{a_{3}}{a_{2}}\right) ^{\frac{3}{2}}-1%
\right] \right\} ,
\label{gwen}
\end{eqnarray}
\\
and
\\
\begin{eqnarray}
S_{g}(\eta _{f0} &<&\eta <\eta _{f1})=\frac{4\pi c^{3}}{3}H_{3}^{-3}A\left(
\frac{a_{3}}{a(\eta )}\right) ^{-3/l}\frac{1}{16\pi ^{3}c^{3}}\left( \frac{%
a_{1}}{a_{3}}\right) ^{3}H_{1}^{3} \nonumber \\
&&\times \left[ \left( \frac{a_{2}}{a_{1}}\right) ^{\frac{1}{2}}\left( \frac{%
a_{3}}{a_{1}}\right) ^{\frac{1}{2}}\left( \frac{a_{3}}{a(\eta )}\right) ^{-%
\frac{1}{l}}-1\right] .
\end{eqnarray}

The RGWs entropy is a decreasing function of the scale factor and
consequently of conformal time. The next section explores whether
this entropy descent can be compensated by an increase of the
entropy of the other contributors, namely, matter and horizon.

\section{The generalized second law}
According the generalized second law of gravitational thermodynamics the entropy of the horizon
plus its surroundings (in our case, the entropy in the volume enclosed by the horizon)
cannot decrease. Consequently, we must evaluate the total entropy to see how
it evolves during the present dark energy era.

The entropy of the horizon (proportional to its area, $4\pi c^{2}\, H^{-2}$),
\\
\begin{equation}
S_{H}=\frac{k_{B}c^{2}\pi }{\ell_{Pl}^{2}}H^{-2}(\eta )=\frac{k_{B}c^{2}\pi }{%
\ell_{Pl}^{2}}H_{3}^{-2}\left( \frac{a_{3}}{a(\eta )}\right) ^{-2-\frac{2}{l}},
\end{equation}
\\
increases with expansion. (Bear in mind that $l < -1$, i.e., we are assuming
that the dark energy behind the acceleration is not of ``phantom" type
\cite{phantom}).

We must also consider the non-relativistic matter fluid. Assuming the latter
consists in particles of mass $m$ and that each of them
contributes $k_{B}$ to the matter entropy, we get
\\
\begin{equation}
S_{m}=k_{B}\frac{\rho _{m}}{m}\frac{4\pi c^{3}}{3}H^{-3}=k_{B}\frac{c^{3}}{%
2mG}H_{3}^{-1}\left( \frac{a_{3}}{a(\eta )}\right) ^{-\frac{3}{l}},
\label{sm}
\end{equation}
\\
for the entropy of the non-relativistic fluid. Here, we made use of the conservation equation
$\rho _{m}(\eta )=\rho _{m}(\eta_{3})\left( \frac{a_{3}}{a(\eta )}\right) ^{3}=\frac{3}{8\pi G}%
H_{3}^{2}\left( \frac{a_{3}}{a(\eta )}\right) ^{3}$   with $\rho_{m}$ the energy density
of matter. From (\ref{sm}), it is apparent that $S_{m}$ decreases with expansion.

In virtue of the above equations, the GSL,
$ S ^{\prime}_{m} + S ^{\prime}_{g}+S ^{\prime}_{H}\geqslant 0$,
where the prime indicates derivation with respect to $\eta$,
can be written as
\\
\[
\frac{3}{l}\left[ \frac{c^{3}}{2mG}H_{3}^{-1}+\frac{4\pi c^{3}}{3}%
AH_{3}^{-3}\left( \frac{a_{2}}{a_{3}}\right) ^{3}n_{g}(\eta _{2})\right]
+\left( 2+\frac{2}{l}\right) \frac{c^{2}\pi }{l_{Pl}^{2}}%
H_{3}^{-2}\left( \frac{a_{3}}{a(\eta )}\right) ^{-2+\frac{1}{l}}
\]
\begin{equation}
+\frac{3}{l}\frac{A}{576\pi ^{4}}\left( \frac{a_{1}}{a_{2}}\right)
^{2}\left( \frac{a_{2}}{a_{3}}\right) ^{3}H_{1}^{3}H_{3}^{-3}\left[ 2\left(
\frac{a_{3}}{a(\eta )}\right) ^{-\frac{3}{l}}\left( \frac{a_{3}}{a_{2}}%
\right) ^{\frac{3}{2}}-1\right] \geqslant 0,
\label{cond1}
\end{equation}
\\
for $\frac{a(\eta )}{a_{3}}<\left( \frac{a_{3}}{a_{2}}\right) ^{-\frac{l}{2}}$,
and
\\
\[
\frac{3}{l}\frac{c^{3}}{2mG}H_{3}^{-1}+\left( 2+\frac{2}{l}\right)
\frac{\pi c^{2}}{l_{Pl}^{2}}H_{3}^{-2}\left( \frac{a_{3}}{a(\eta )}%
\right) ^{-2+\frac{1}{l}}
\]
\begin{equation}
+\frac{3}{l}\frac{A}{12\pi ^{2}}\left( \frac{a_{1}}{a_{3}}\right)
^{3}H_{1}^{3}H_{3}^{-3}\left[ \frac{4}{3}\left( \frac{a_{2}}{a_{1}}\right) ^{%
\frac{1}{2}}\left( \frac{a_{3}}{a_{1}}\right) ^{\frac{1}{2}}\left( \frac{%
a_{3}}{a(\eta )}\right) ^{-\frac{1}{l}}-1\right] \geqslant 0,
\label{cond2}
\end{equation}
\\
for $\frac{a(\eta )}{a_{3}}>\left( \frac{a_{3}}{a_{2}}\right) ^{-\frac{l}{2}}$.

For $l<-1$, both conditions are of the type
$f\left( \frac{a(\eta )}{a_{3}}\right) \geqslant 0$, $\, \,  f$ being an increasing
function of $a(\eta)$. Therefore,  if the condition holds true at
the beginning of the dark energy era, $\eta = \eta _{3}$, it will
hold for $\eta > \eta_{3}$.

By setting $a(\eta ) = a_{3}$ in Eq.(\ref{cond1}), a restriction
over the unknown constant of proportionality $A$ follows
\\
\begin{equation}
A\leqslant \frac{-\left( 2l+2\right) \frac{k_{B}c^{2}\pi }{3l_{Pl}^{2}}%
H_{3}^{{}}-\frac{k_{B}c^{3}}{2mG}H_{3}^{2}}{\frac{4\pi c^{3}}{3}\left( \frac{%
a_{2}}{a_{3}}\right) ^{3}n_{g}(\eta _{2})+\frac{1}{576\pi ^{4}}\left( \frac{%
a_{1}}{a_{2}}\right) ^{2}\left( \frac{a_{2}}{a_{3}}\right) ^{3}H_{1}^{3}%
\left[ 2\left( \frac{a_{3}}{a_{2}}\right) ^{\frac{3}{2}}-1\right] }\, ,
\label{bound}
\end{equation}
\\
implying that for the GSL to be satisfied the above upper bound must be met.
In this case the event horizon soon comes to dominate the total entropy
and steadily augments with expansion. So, even though the entropy of
matter and RGWs within the horizon decrease during the present dark
energy era, the GSL is preserved provided Eq. (\ref{bound}) holds.
Note that since Nesteruk and Ottewill left the constant $A$
unspecified \cite{nesteruk} restriction (\ref{bound}) turns
to be all the more important: it is the only knowledge
we have about how much big $A$ may be.

Obviously, our conclusions hang on the expression adopted for the entropy density
of the gravitational waves. Here we have chosen (\ref{ng}) since, on the one hand,
it is the simplest one based on particle production in curved spacetimes \cite{curved},
and on the other hand, $s_{g}$ cannot fail to be an increasing function of $n_{g}$.
We believe, that any sensible expression for $s_{g}$ should not run into conflict
with the GSL, and that the latter may impose restrictions on the parameters
entering the former.

As mentioned above, we have left aside models of late acceleration driven by dark energy
of ``phantom type" (i.e., $ -1 < l <0$) \cite{phantom}. In this case, owing to the fact
that the dominant energy condition is violated, the event horizon decreases with
expansion whereby it is rather doubtful that the GSL may be satisfied at all.  We
shall focus our attention on this issue in a future research.

\acknowledgments{G.I. acknowledges support from the ``Programa de Formaci\'{o}
d'Investigadors de la UAB". This work was partially supported by the Spanish
Ministry of Science and Technology under grant BFM2003-06033.}

\end{document}